\documentclass[10pt,aps,prb,twocolumn,amssymb,footinbib,letterpaper,superscriptaddress]{revtex4-1}
\usepackage{graphicx}
\usepackage{color}
\usepackage{dcolumn} 
\usepackage{bm}      
\usepackage{array}   
\usepackage{amssymb}
\usepackage{amsfonts}
\usepackage{amsmath}
\usepackage[colorlinks,citecolor=red]{hyperref}
\newcommand{\Eex}{ E_{exact} }
\newcommand{\Evar}{ E_{var} }

\newcommand{\sysa}{\mbox{square}\,\,\mbox{10 site}}
\newcommand{\sysc}{{\mbox{square}\,\,4\times 4}}
\newcommand{\sysf}{{\mbox{kagome}\,\,2\times 2\times 3}}
\newcommand{\red}[1]{{\color{red} #1}}

\newcolumntype{.}[1]{D{.}{.}{#1}}

\newcommand{\be}{\begin{equation}}
\newcommand{\ee}{\end{equation}}
\newcommand{\ba}{\begin{eqnarray}}
\newcommand{\ea}{\end{eqnarray}}

\newcommand{\upa}{\uparrow}
\newcommand{\dna}{\downarrow}

\newcommand{\COMMENTED}[1]{}

\newcommand{\ket}[1]{|#1\rangle}

\newcommand{\ca}[2]{{\hat c}_{#1 #2}^\dagger}
\newcommand{\de}[2]{\hat{c}_{#1 #2}^{\phantom{\dagger}}}

\newcommand{\ob}[1]{{\langle #1\rangle}}

\newcommand{\hH}{{\hat{H}}}

\newcommand{\hK}{{\hat{K}}}

\newcommand{\hn}{{\hat{n}}}

\begin{document}

\title{Auxiliary-field based trial wave functions in quantum Monte Carlo calculations}

\author{Chia-Chen Chang}
\affiliation{Department of Physics, University of California Davis, CA 95616, USA}
\author{Brenda M. Rubenstein}
\affiliation{Quantum Simulations Group, Lawrence Livermore National Laboratory, Livermore, CA 94550, USA}
\affiliation{Department of Chemistry, Brown University, Providence, RI, 02912, USA}
\author{Miguel A. Morales}  
\affiliation{EOS and Materials Theory Group, Lawrence Livermore National Laboratory, Livermore, CA 94550, USA}

\begin{abstract}

Quantum Monte Carlo (QMC) algorithms have long relied on Jastrow factors to incorporate dynamic 
correlation into trial wave functions. While Jastrow-type wave functions have been 
widely employed in real-space algorithms, they have seen limited use in second-quantized QMC methods,
particularly in projection methods that involve a stochastic evolution of the wave function in imaginary time.
Here we propose a scheme for generating Jastrow-type correlated trial wave functions for auxiliary-field
QMC methods. The method is based on decoupling the two-body Jastrow into one-body projectors coupled 
to auxiliary fields, which then operate on a single determinant to produce a multi-determinant trial 
wave function. We demonstrate that intelligent sampling of the most significant determinants in this 
expansion can produce compact trial wave functions that reduce errors in the calculated energies.
Our technique may be readily generalized to accommodate a wide range of two-body 
Jastrow factors and applied to a variety of model and chemical systems. 
\end{abstract}

\maketitle

\section{Introduction}

The development of predictive quantum simulation methods is one of the 
foremost challenges in the fields of quantum chemistry and condensed matter physics.
One step toward being able to accurately predict the properties of a variety of complex molecules 
and solids is to develop improved variational trial wave functions\cite{Toulouse:2015,Needs:2010} 
for projection quantum Monte Carlo methods, such as 
diffusion Monte Carlo (DMC).\cite{Toulouse:2015,Needs:2010,Umrigar:1988,Umrigar:2005,Foulkes:2001} 
In these methods, the trial wave function serves not only as an importance function to drive the 
sampling of configurations, but also as a constraint used to suppress the development 
of the sign/phase problems. Accurate variational wave functions are therefore pivotal for guaranteeing 
convergence to the correct ground state energy with minimal bias and for improving
the efficiency of simulations.\cite{Huang:1997,Toulouse:2008,Petruzielo:2012} 
This is especially true for strongly correlated systems, for which non-interacting or mean field trial 
wave functions are known to yield substantial statistical and systematic errors.\cite{Rubenstein:2012} 
Developing more accurate variational wave functions is thus a crucial step toward being able to properly 
model many technologically important, yet theoretically challenging materials, such as 
high-$T_c$ superconductors, the lanthanides and actinides.

One route toward more accurate variational wave functions for larger, multidimensional systems has  
been to develop more sophisticated variational ansatzes, 
most of which explicitly include some amount of correlation. Such forms include antisymmetric geminal product 
(AGP),\cite{Casula:2003, Casula:2004, Neuscamman:2012, Neuscamman:2013, Neuscamman:20132, Neuscamman:2015} 
Bardeen-Cooper-Schrieffer (BCS),\cite{Guerrero:2000ws,Carlson:2011in,Guerrero:1999vc} 
Pfaffian,\cite{Bajdich:2006,Bajdich:2008} and matrix product state (MPS) wave functions.\cite{Wouters:2014} All of these forms 
have a long history of being used in calculations performed at the variational level, but have assumed
a more limited role in projector QMC calculations.
A second path toward more accurate variational states is to create such wave functions by applying a 
physically-inspired projection operator onto a trial wave function. For years, the DMC 
community has generated trial wave functions using Jastrow factors containing one-, two-, and/or three-body 
terms that, among other things, provide a compact way of reinforcing cusp 
conditions.\cite{Huang:1997,Toulouse:2015,Needs:2010,Foulkes:2001,Clark:2011,Morales:2012,Clay:2015} 
These Slater-Jastrow wave functions and the advent of new techniques for variationally optimizing 
them\cite{Umrigar:2007, Toulouse:2007} have greatly expanded the fidelity and 
reach of this method in recent years. Symmetry-projected wave functions have also been shown to recover 
substantial portions of the correlation energy at the variational level and to considerably reduce the 
statistical noise observed in auxiliary-field quantum Monte Carlo (AFQMC) calculations when used as trial wave 
functions.\cite{Shi:2014bw,Shi:2013jj,RodriguezGuzman:2013bwa, RodriguezGuzman:2012,JimenezHoyos:2013} 
Even more sophisticated projectors could be imagined, but key to unlocking their potential is the ability 
to apply and evaluate them in an efficient manner in the framework of AFQMC.

In this work, we propose a scheme to create strongly correlated variational/trial wave functions  
by exploiting the Hubbard-Stratonovich (HS) transformation, commonly 
used to decouple the Coulomb term in AFQMC simulations, to decouple two-body projection 
operators.\cite{Hirsch:1985} Based upon this scheme, we generate Slater-Jastrow trial states for use in 
second-quantized projector QMC methods, thus 
extending the benefits of the Jastrow wave function beyond the realm of first-quantized techniques. 
Within our method, the exact form of the Slater-Jastrow wave function yields a multi-determinant expansion 
whose size scales exponentially with the system size. We therefore explore a few techniques that allow us to 
generate representations that quickly converge to the exact variational energy using but a fraction of the 
total number of determinants. In this paper, we use the one-band Hubbard model and the Gutzwiller projector, 
the simplest form of a Jastrow
factor, to demonstrate our methodology. Nevertheless, the method we propose is completely general and 
can be extended to more sophisticated wave function forms and systems.

\section{Method}

\begin{figure}[b]
\includegraphics[scale=0.38]{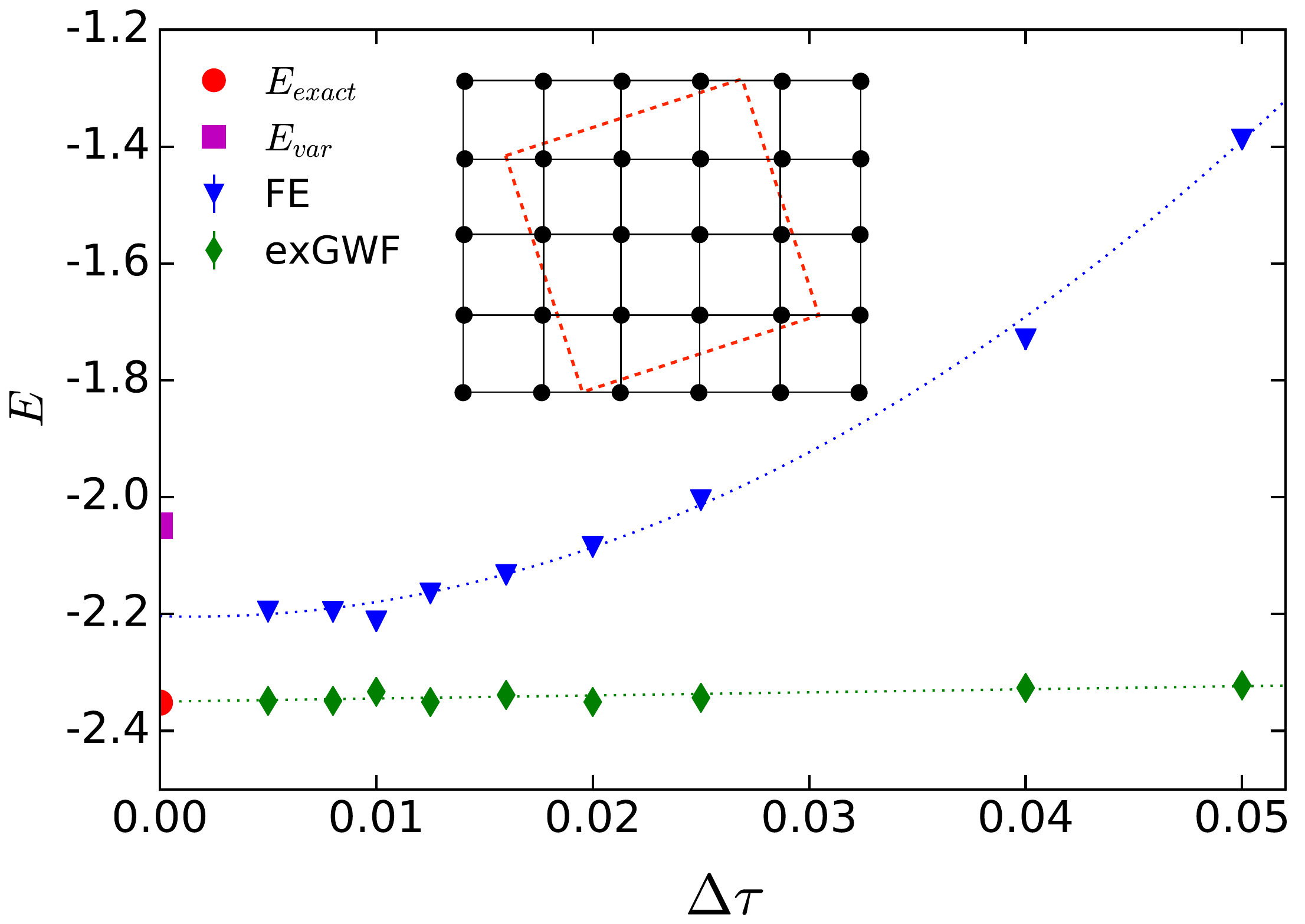}
\caption{Extrapolation of the Trotter approximation error to $\Delta \tau=0$. 
The CPMC energy is plotted as a function of the time step $\Delta\tau$ for the 
half-filled 10-site square lattice at $U=20$. Dotted lines are second order 
polynomial fits to the data.
The exact energy $E_{exact}$ (red circle) and exGWF variational energy $E_{var}$
(purple square) are also included for reference.
The (red) dashed square in the inset indicates the geometry of the 10-site square
lattice.
}
\label{fig:TrotterError}
\end{figure}

\subsection{The Gutzwiller wave function}

We choose to demonstrate our scheme on the modified Gutzwiller wave function (GWF)
defined as\cite{Otsuka:1992wa,Yanagisawa:1998}
\begin{align}
  \ket{\Psi_G} &= e^{-h\hK}\hat{\cal P}_G \ket{\Phi} =
  e^{-h\hK} e^{-\beta\sum_i \hn_{i\upa}\hn_{i\dna}} \ket{\Phi}.
  \label{eq:PsiG}
\end{align} 
Here, $\ket{\Phi}=\ket{\Phi_\upa}\otimes\ket{\Phi_\dna}$ denotes a single Slater 
determinant, such as the free-electron or Hartree-Fock wave function. $\beta > 0$ 
is a variational parameter. The projector $\hat{\cal P}_G$,\cite{Gutzwiller:1963ug} 
which is the simplest Jastrow correlator, introduces correlations among electrons by 
suppressing doubly occupied configurations in $\ket{\Phi}$. 
$\hK$ is a one-body operator often chosen to be
the kinetic energy term of the Hamiltonian, and $h$ is also a variational parameter.
It is shown that the projector $e^{-h\hK}$ enhances kinetic exchange, and can improve 
the variational energy \red{of} the Hubbard model.\cite{Otsuka:1992wa,Yanagisawa:1998}
For simplicity, we still refer to the state given by Eq.~(\ref{eq:PsiG}) as the Gutzwiller
wave function.

The two-body nature of ${\cal P}_G$ hinders the direct application of $\ket{\Psi_G}$ 
in QMC simulations. Nonetheless, using the discrete HS
transformation,\cite{Hirsch:1985} the projector can be decoupled as follows  
\be
  \ket{\Psi_G} = 
  \sum_{\{s_i\}} e^{-h\hK} \prod_i\,e^{(-\frac \beta 2 + \gamma s_i) \hn_{i\upa}}
                                    e^{(-\frac \beta 2 - \gamma s_i) \hn_{i\dna}}\ket{\Phi},
  \label{eq:PsiG_decoupled}
\ee
where $\cosh\gamma = e^{\beta/2}$, and $s_i=\pm 1$ is the auxiliary field on the $i$-th 
site. The Gutzwiller wave function produced after decoupling may be viewed as a finite 
sum over determinants, each of which is a function of a discrete set of HS fields 
$(s_1, s_2, s_3, \ldots)$. We refer to this wave function (Eq.~(\ref{eq:PsiG_decoupled})) 
as the exact GWF (exGWF). 

For a given system size $L$ and filling $\rho=(N_\upa+N_\dna)/L$ (where $N_\upa$ and $N_\dna$ 
represent the number of spin-up and spin-down electrons), we optimize the variational energy 
$\Evar=\ob{\Psi_G|\hH|\Psi_G}/\ob{\Psi_G|\Psi_G}$ as a function of $(\beta, h)$.
We have verified that our optimized $(\beta,h)$ are consistent with those reported in 
Ref.~\onlinecite{Otsuka:1992wa} for the half-filled Hubbard model in one and two dimensions.

\subsection{The Hubbard model and the Constrained-Path Monte Carlo algorithm}

To showcase the GWF, we study the ground state of the one-band repulsive Hubbard model 
in two dimensions using the constrained-path Monte Carlo (CPMC) technique.\cite{Zhang1997,Zhang2003} 
The system is defined by the Hamiltonian
\begin{align}
 \hH &= -t\sum_{i,\sigma} \left( \ca{i}{\sigma}\de{i+1,}{\sigma} + \ca{i+1,}{\sigma}\de{i}{\sigma}
                         \right)
       + U\sum_i \hn_{i\upa}\hn_{i\dna}. 
     \label{eq:HubbardModel}
\end{align}
The parameters $t$ and $U$ represent the hopping amplitude and on-site repulsion,
respectively. $\ca{i}{\sigma}$ ($\de{i}{\sigma}$) creates (destroys) an electron with spin
$\sigma=\upa,\,\dna$ at site $i$, and $\hn_{i\sigma}$ is the number operator for a spin-$\sigma$
electron. 

The CPMC algorithm is an AFQMC method that works in the second-quantized 
framework. For a detailed discussion of the CPMC method and benchmark results, we refer readers
to Ref.~\onlinecite{Zhang1995,Zhang1997,Zhang2003}. Here we note that CPMC eliminates the 
sign problem much as the fixed-node approximation does in DMC by rejecting random 
walkers that have negative overlaps with the trial wave function. 
We use $t$ as the unit of energy and set $t=1$ throughout this work.

\section{Results and Discussion}

In order to demonstrate the benefits of using a Slater-Jastrow wave function, we first examine how 
the quality of the trial wave function affects the magnitude of the systematic Trotter factorization 
error. To do so, we compare the ground state energies obtained using the free electron (FE) and exGWF 
trial wave functions for various time steps $\Delta \tau$ for the half-filled 10-site 2D Hubbard model 
at $U=20$ under periodic boundary conditions.
Although quantum Monte Carlo does not exhibit the sign problem at this 
filling, we deliberately apply the constrained-path approximation\cite{Zhang1997} in the simulations so 
that we can gauge how the bias and errors that result from the constrained-path approximation vary with the 
quality of trial states.
In both sets of calculations, we have utilized the second order Trotter break-up formula for 
the propagators. Fig.~\ref{fig:TrotterError} compares the correction of the Trotter error obtained by 
extrapolating the CPMC energy to the limit $\Delta\tau\rightarrow 0$. 
As illustrated in Fig.~\ref{fig:TrotterError}, the FE case has a strong dependence on $\Delta\tau$, and 
the extrapolated energy is off by 6.2\%.
In contrast, the energies are not only more accurate, but have a much weaker dependence on 
$\Delta\tau$ when the exGWF is used Similar conclusions may be drawn from other simulation 
results (e.g. Fig.~\ref{fig:TrotterErrorKagome} in Appendix \ref{sec:appendix}).

Next, we compare the fully extrapolated CPMC ground state energy obtained using a FE trial state with that 
obtained using the exGWF. We again consider the half-filled 2D Hubbard model on the 10-site square lattice 
with periodic boundary conditions, and retain the constrained-path approximation in order to gauge the
effects of the different trial wave functions.

\begin{table}[b]
\begin{ruledtabular}
\begin{tabular}{r.{2.5}.{2.5}.{2.4}.{2.5}}
  \multicolumn{1}{c}{$U$} &  \multicolumn{1}{c}{$\Eex$}   &  \multicolumn{1}{c}{$\Evar$}   
                          &  \multicolumn{1}{c}{CPMC$+$FE}   &  \multicolumn{1}{c}{CPMC$+$exGWF} \\
  \hline   
  10  &  -4.2821 & -3.9708  &  -4.1285(31) &  -4.2831(7)   \\                    
  12  &  -3.6872 & -3.3485  &  -3.5316(12) &  -3.6873(9)  \\
  16  &  -2.8771 & -2.5444  &  -2.7259(17) &  -2.8790(19)  \\
  20  &  -2.3517 & -2.0488  &  -2.2037(18) &  -2.3498(12)  \\
\end{tabular}
\end{ruledtabular}
\caption{Ground state energies of the half-filled 2D Hubbard model on the 10-site square
lattice. $\Eex$ denotes exact diagonalization results. $\Evar$ is the optimized variational energy of 
exGWF. The last two columns show the CPMC energies with free electron (FE) and optimized exGWF trial 
wave functions respectively. Numbers in parenthesis are statistical errors.}
\label{tbl:exGWF}
\end{table}

\begin{figure}[b]
\includegraphics[scale=0.42]{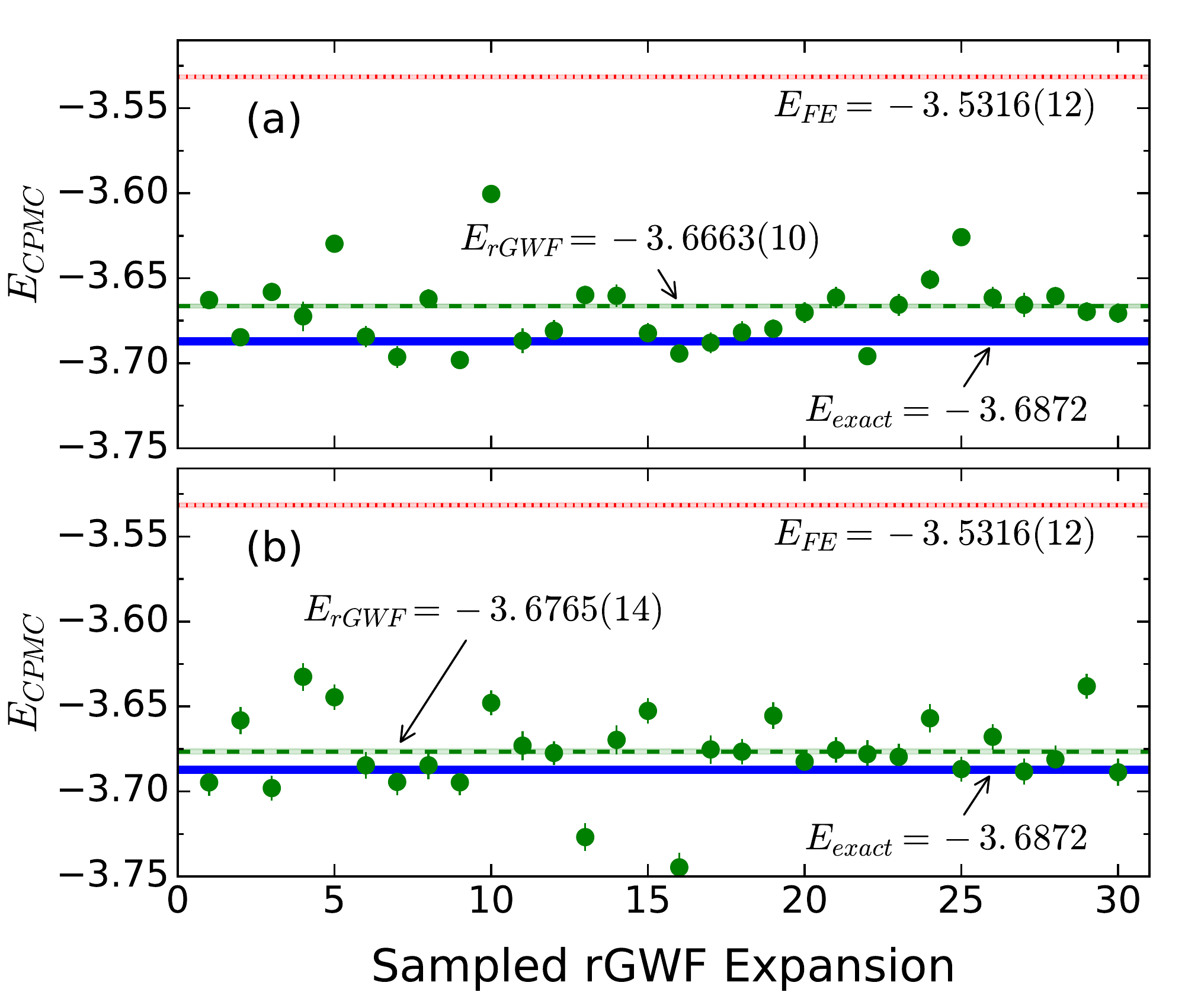}
\caption{CPMC energies for the half-filled 10-site 2D Hubbard model at $U=12$ obtained using a rGWF 
as the trial state. The rGWF consists of 100 and 800 determinants in panels (a) and (b) respectively. 
In each figure, a filled (green) dot represents a single simulation using a rGWF. The dashed (green) line 
represents the average obtained by averaging over 30 simulations using different rGWFs. The width of the 
shaded area is twice the estimated error.
FE and exact diagonalization results are plotted as dotted (red) and solid (blue) lines 
respectively.
}
\label{fig:RandomSampling}
\end{figure}

The results of this comparison are listed in Table \ref{tbl:exGWF}. The deviation between the exact and FE 
trial wave function results generally grows with $U$, and can be as large as $6.2\%$ at $U=20$. The exGWF data, 
on the other hand, are in excellent agreement with the exact energies regardless of $U$. As manifest in 
Table \ref{tbl:sfGWF}, the same conclusion may be drawn for other half-filled systems.
While Table \ref{tbl:exGWF} illustrates that the exGWF is an accurate trial wave function for the system 
examined, the exGWF quickly becomes computationally intractable because the number of determinants in the 
expansion in Eq.~(\ref{eq:PsiG_decoupled}) scales exponentially with $L$. In order to reduce the computational 
cost, we propose several compact representations of the exGWF and discuss their performance below.

\underline{\it Method I}\,: Using Monte Carlo sampling, we construct a representation of 
the exGWF by randomly choosing determinants from the $2^L$ states of the full exGWF expansion. The wave 
functions constructed in this manner will be called randomly-sampled GWFs (rGWFs).
Fig.~\ref{fig:RandomSampling} illustrates results obtained from using 30 independent rGWF samples for 
the half-filled 10-site 2D Hubbard at $U=12$. In panels (a) and (b), each rGWF consists of 100 and 800 
determinants respectively. The final CPMC energy is computed by averaging the 30 simulations in each case.
Using 100-determinant rGWFs as trial states, the averaged CPMC ground state energy is about $0.56\%$ away
from the exact result, which compares favorably against the $4.22\%$ deviation of the FE result.
By increasing the sample size to 800 determinants, the deviation is reduced to $0.29\%$. 
We note that the average overlap between the rGWF and exGWF is $34\%$ and $72\%$ for the 100- and 800-determinant
respectively. The higher overlap explains the improvement seen in the 800-determinant data, as more terms 
are involved in the sampling process.

Although each rGWF contains a subset of terms from the exGWF, the comparisons demonstrate that the approach
could still capture the essential physics of the exact Gutzwiller wave function. There are two factors that
can affect the accuracy and computational cost of the rGWF: the number of determinants in a given sample, 
and the total number of independent rGWF samples. As the system size $L$ and interaction strength $U$ are 
increased, we expect the number of determinants needed to achieve the same level of 
accuracy as the exGWF to also increase for a fixed number of independent samples.

\begin{figure}[t]
\includegraphics[scale=0.39]{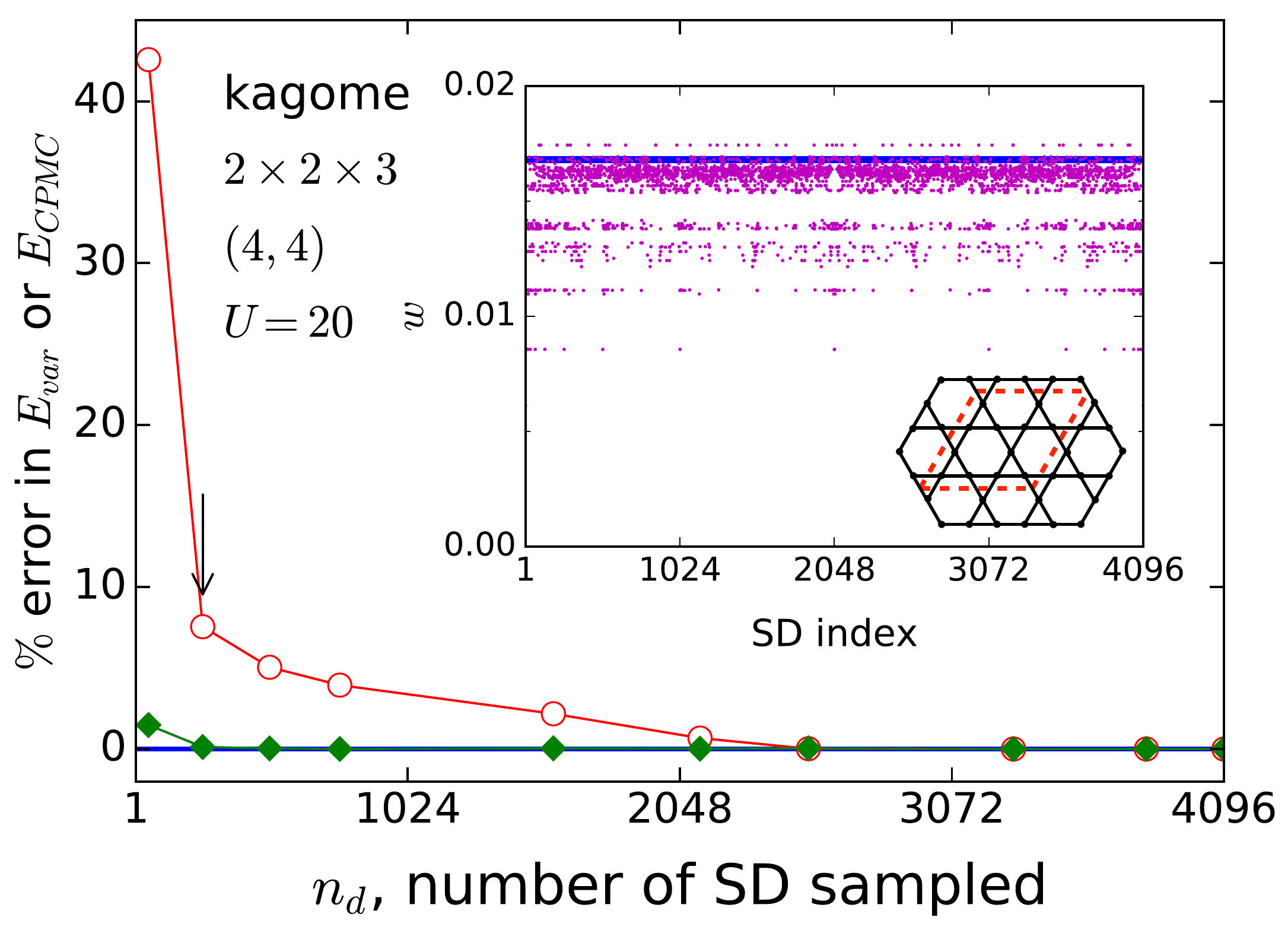}
\caption{Circles (red): Relative error in the variational energy of the ``importance-sampled'' 
GWF for the $2\times 2\times 3$ kagome lattice doped with 4 holes at $U=20$. The reference is 
the exGWF variational energy. The horizontal axis is the number of determinants $n_d$ sampled. 
Diamonds (green): Relative error of the CPMC results with respect to the exact energy for the 
same system.
Inset: Weight distribution of determinants in the exGWF. The solid (blue) 
horizontal line indicates the cutoff $w_c=0.0168$.  
The resulting variational and CPMC energies of this state are highlighted by the vertical arrow 
in the main panel. 
The (red) dashed parallelogram indicates the simulation cell of the 12-site kagome lattice.
}
\label{fig:CoeffKagome}
\end{figure}

\underline{\it Method II}\,: In the rGWF approach, because the determinants (and their corresponding 
HS field configurations) are selected randomly, one clear way of reducing the effort needed to sample 
rGWFs is to select the determinants more intelligently. This is precisely what motivates importance 
sampling in efficient Monte Carlo algorithms. To make progress, we proceed as follows.
Let $\ket{\phi_i}$ $(i=1,2,\ldots, 2^L)$ denote the determinants in the expansion 
Eq.~(\ref{eq:PsiG_decoupled}). We construct a Hamiltonian matrix 
$[M]_{ij} = \ob{\phi_i|\hH|\phi_j} / \ob{\phi_i|\phi_j}$ using the non-orthogonal determinants 
$\{\ket{\phi_i}\}$. After diagonalizing $M$, we interpret the eigenvector of the {\it lowest} 
eigenvalue of the matrix as the weight $w$ of the determinant $\ket{\phi_i}$. 

The inset of Fig.~\ref{fig:CoeffKagome} shows one example of the distribution of the determinant weights 
for the $2\times 2\times 3$ kagome lattice doped with 4 holes at $U=20$.
Based on the information in $w$, we construct our trial wave functions by linearly combining determinants 
with weights satisfying $w > w_c$, where $w_c$ is a cutoff, and study their variational energy as a 
function of the number of determinants $n_d$ (hence $w_c$) retained. The empty (red) circles in the main 
panel of Fig.~\ref{fig:CoeffKagome} depict the results for the doped 12-site kagome lattice system.
As the curve indicates, the variational energy quickly converges with the number of ``important'' (i.e., large 
weight) determinants included in the wave function. For instance, in the main panel of Fig.~\ref{fig:CoeffKagome}, 
the vertical arrow indicates a state containing 252 determinants, corresponding to the cut-off $w_c=0.0168$. 
This state contains $\sim 6.2\%$ of the total determinants, and has a $87.9\%$ overlap with the exGWF. 
Its energy is $92.4\%$ of the exact GWF variational energy. 

These results indicate that, as long as importance sampling is employed, it is possible to construct a trial 
wave function much reduced in size which still recovers a sizable fraction of the variational energy.
The same trend carries over to CPMC calculations using these same trial wave functions. In fact, as shown in 
the main panel of Fig.~\ref{fig:CoeffKagome}, the CPMC energies converge faster than the variational energies do.
For example, using the 252-determinant trial state indicated by the vertical arrow, the resulting CPMC result
is $0.12\%$ away from the exact energy. Using the next exGWF representation (which contains 504 determinants),
the deviation is further reduced to $0.044\%$. Therefore, by properly sampling the most important determinants, 
one is able to create an accurate representation of the exact Gutzwiller wave function that is also 
computationally tractable.

\begin{figure}
\includegraphics[scale=0.43]{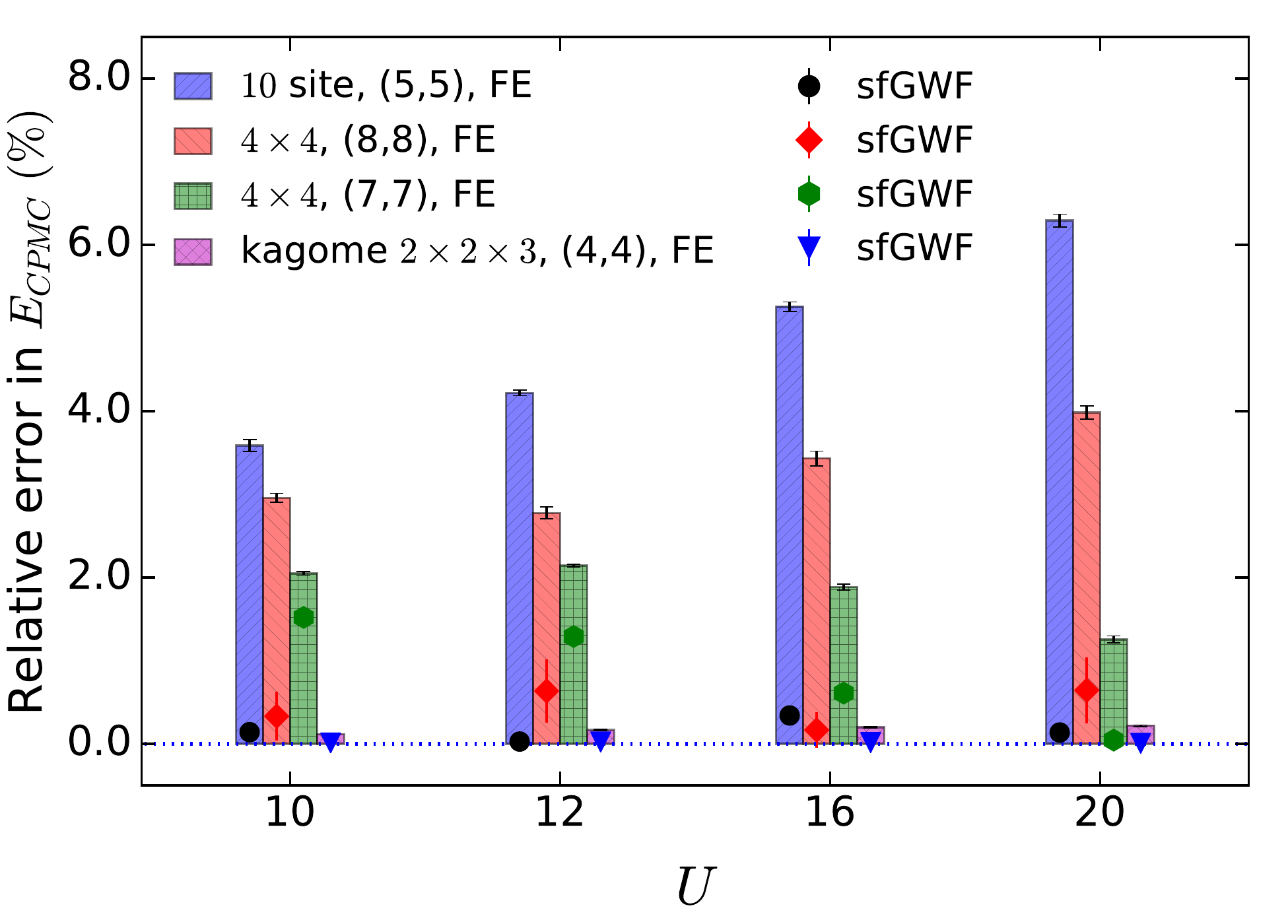}
\caption{Relative error (absolute value) in the CPMC energy with FE and sfGWF trial wave functions for 
half-filled and doped 2D Hubbard models. Detailed comparisons for the doped kagome lattice can be found in 
Appendix \ref{sec:appendix}.
}
\label{fig:sfGWF}
\end{figure}

\begin{figure}
\includegraphics[scale=0.32]{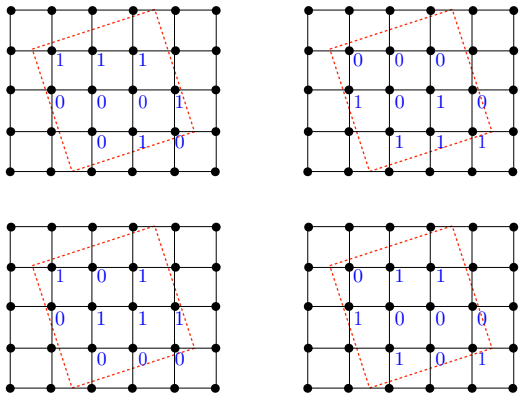}
\caption{Examples of Hubbard-Stratonovich field configurations that are connected through 
translation operation in the $y$-direction. Determinants generated by these fields have 
essentially the same weight after the diagonalization procedure discussed in Method III.
}
\label{fig:HSfieldSymmetry}
\end{figure}

\underline{\it Method III}\,: In order to gain insight into the distribution of weights, we took a closer 
look at the determinants' 
corresponding HS field configurations. Let 1 and 0 denote the field values $+1$ and $-1$ respectively. 
Drawing upon the half-filled 10-site square lattice case as an example, we make the following observations. 
Firstly, the field configurations of many of the important determinants are all permutations of the 
configuration $(1111100000)$, which has an equal number of $+1$ and $-1$ fields. 
Secondly, the field structures with nearly degenerate weights (as indicated by the ``band''-like structure 
in the inset of Fig.~\ref{fig:CoeffKagome}) are related via translational symmetry. For instance, determinants 
generated from configurations $(1110001010)$ and $(0001010111)$ have almost identical weights while these
two configurations are related via a translation in the $y$-direction by one lattice constant, c.f. 
Fig.~\ref{fig:HSfieldSymmetry}.

Based on this idea, we generate a trial wave function (denoted as sfGWF) using only the HS field configuration 
$(1111100000)$ and its permutations. The resulting sfGWF has 
$C^{10}_5=252$ determinants and a surprisingly large overlap $99.98\%$ with the exGWF. For the same system,
we observe the same behavior at other interaction strengths: sfGWFs constructed in this manner ubiquitously 
have almost unity overlap with the corresponding exGWF.

Encouraged by these observations at half-filling, we subsequently considered the $2\times 2\times 3$ kagome
lattice doped with 4 holes. This is a closed-shell filling under periodic boundary conditions. The coefficients 
from diagonalizing the matrix $[M]_{ij}$ indicate that highly-weighted determinants 
have two degenerate HS field configurations: 
\be
 (\underbrace{11\ldots\ldots1}_{(L+h)/2}\underbrace{00\ldots\ldots0}_{(L-h)/2})\hskip.1in   \mbox{and}\hskip.1in
 (\underbrace{11\ldots\ldots1}_{(L-h)/2}\underbrace{00\ldots\ldots0}_{(L+h)/2}), \nonumber
\ee
where $h$ is the number of holes. Because these configurations are degenerate, we adopt one of them to 
construct the sfGWF trial wave functions for our doped system.

Benchmark results of the sfGWFs for the Hubbard model are depicted in Fig.~\ref{fig:sfGWF}. FE trial wave 
function data are also included for comparison. More detailed data can be found in Appendix \ref{sec:appendix}. 
In the case of the half-filled 10-site square lattice and the doped 12-site kagome lattice, the non-interacting 
ground state is closed-shell under periodic boundary conditions. On the $4\times 4$ square lattice, we have
implemented twist boundary conditions in order to have closed-shell free-electron states at the fillings
considered.
In all cases, the CPMC ground state energy is improved when sfGWF is adopted as the trial wave function.
This is particularly true for the 10-site square and 12-site kagome lattice systems where the sfGWF results
are almost exact. 
On the $4\times 4$ square lattice, the deviation in the sfGWF result is typically less than $1\%$ from the 
exact energy for half-filling. In the doped case, the error only becomes smaller at large couplings. 

Before we close the discussion, we would like to make a few remarks regarding the three methods presented 
in this section. Because the exGWF expansion scales exponential with system size, reducing the computational 
cost of using Slater-Jastrow wave functions like the Gutzwiller wave function discussed here is essential to 
making these wave functions practically useful. The ``importance sampling'' scheme (i.e., Method II) 
appears to be the most efficient
approach, at least for the clusters tested. To converge the CPMC energy, the number of dominant determinants
required is only a fraction of the total number of terms $2^L$. Obviously, the diagonalization technique 
is only suitable for small clusters. A full variational approach will be required for large simulation cells
and realistic Hamiltonians in quantum chemistry. This idea will be explored in a future publication.

The computational cost of the proposed sfGWF approach scales as 
$C^L_n$ ($2n$ being the total number of electrons), which compares favorably with the $2^L$ scaling 
of the exGWF, but nevertheless is substantial. The cost may be further reduced if the symmetry 
among degenerate fields is exploited. We also note that the HS fields generated in the sfGWF do not exhaust 
all the highly-weighted determinants. This may be responsible for the slightly larger deviation 
(comparing to exGWF data) observed in Table \ref{tbl:sfGWF} for systems simulated with periodic boundary 
conditions.

Lastly, the comparisons in Fig.~\ref{fig:sfGWF} and Table \ref{tbl:sfGWF} indicate that the best agreement 
between the sfGWF and exact energies is achieved for closed-shell systems with PBCs. For the $4\times 4$ lattice 
cases, while twist boundary 
conditions allow the free-electron state at any filling to be unique (i.e., closed-shell), they nevertheless break 
the $C_4$ rotational symmetry of the lattice. We speculate that the effective magnetic flux resulting 
from twist boundary conditions may be responsible for the relatively large deviations in the sfGWF results because
the decomposition Eq.~(\ref{eq:PsiG_decoupled}) breaks the $SU(2)$ spin symmetry. This speculation is partly 
supported by the fact that, for the $4\times 4$ lattice system, the variational energy is at least $10\%$ higher
than the exact energy and can be as large as $47.8\%$ (half-filled, $U=20$). In contrast, the variational
energy of the sfGWF is quite accurate for the 10-site square and 12-site kagome lattice cases, with the
smallest deviation being only $0.55\%$ (4-hole doped 12-site kagome at $U=10$, c.f., Table \ref{tbl:sfGWF}
in the Appendix).
In addition to the ``spin'' decomposition scheme Eq.~(\ref{eq:PsiG_decoupled}), one could also employ the
charge decomposition which preserves the spin rotation symmetry.\cite{Hirsch:1983} This issue of different 
HS decomposition techniques will be explored in subsequent works.

\section{Summary}
Using the Gutzwiller projected wave function as an example, we have illustrated an auxiliary-field based 
scheme for generating Slater-Jastrow trial wave functions for second-quantized AFQMC simulations. We have 
shown that, by intelligently sampling multi-determinant representations of these wave functions, 
we can produce trial wave functions that recover substantial amounts of both the variational and correlation 
energies. These wave functions decrease CPMC errors when compared with those produced by
traditional AFQMC techniques that rely on single determinant trial wave functions. Although the HS field 
structure is unique to the discrete HS transformation adopted in this work, the 
results presented shed light on how to develop a more efficient sampling scheme for more general 
Jastrow-type wave functions, paving the way toward more accurate AFQMC simulations of not only 
strongly-correlated model systems, but of molecules and solid-state materials as well. 
This work was performed under the auspices of the U.S. Department of Energy by Lawrence 
Livermore National Laboratory under Contract DE-AC52-07NA27344, 15-ERD-013. 
The authors would like to thank Hao Shi for providing exact diagonalization results for the
$4\times 4$ square lattice.

\appendix
\section{Benchmarking the sfGWF}\label{sec:appendix}

\begin{figure*}
\includegraphics[scale=0.4]{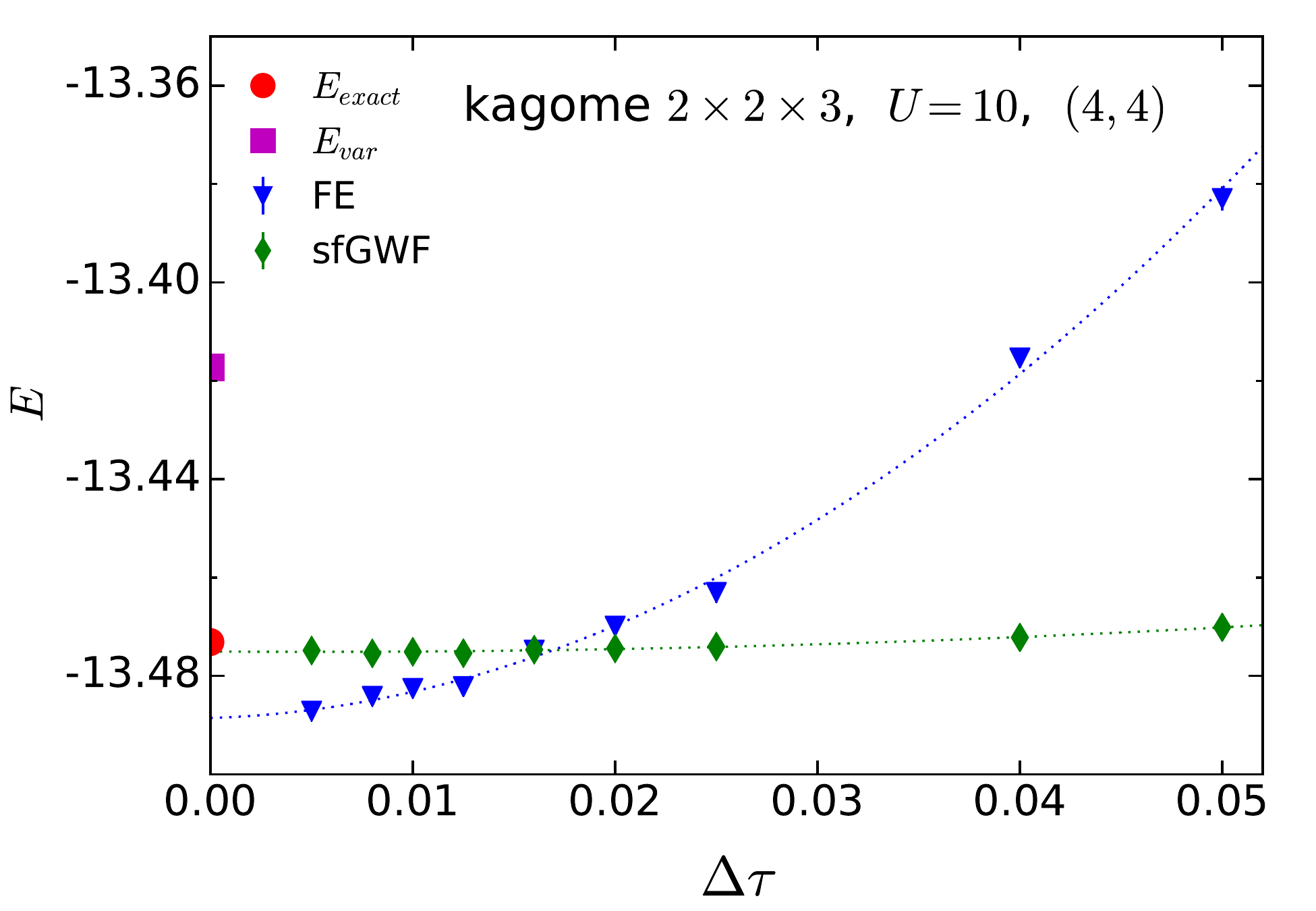}\hspace{4mm}
\includegraphics[scale=0.4]{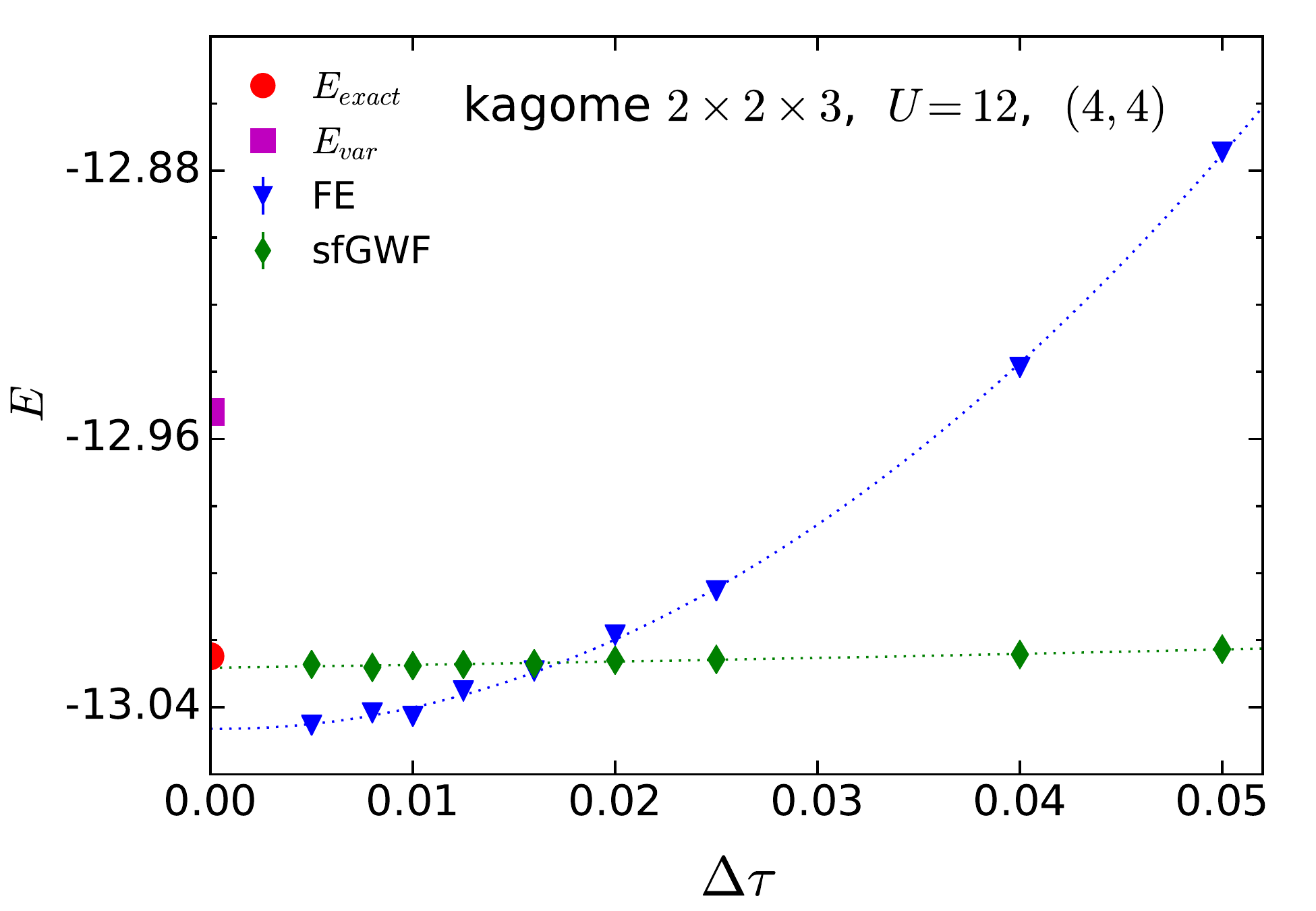}\\
\includegraphics[scale=0.4]{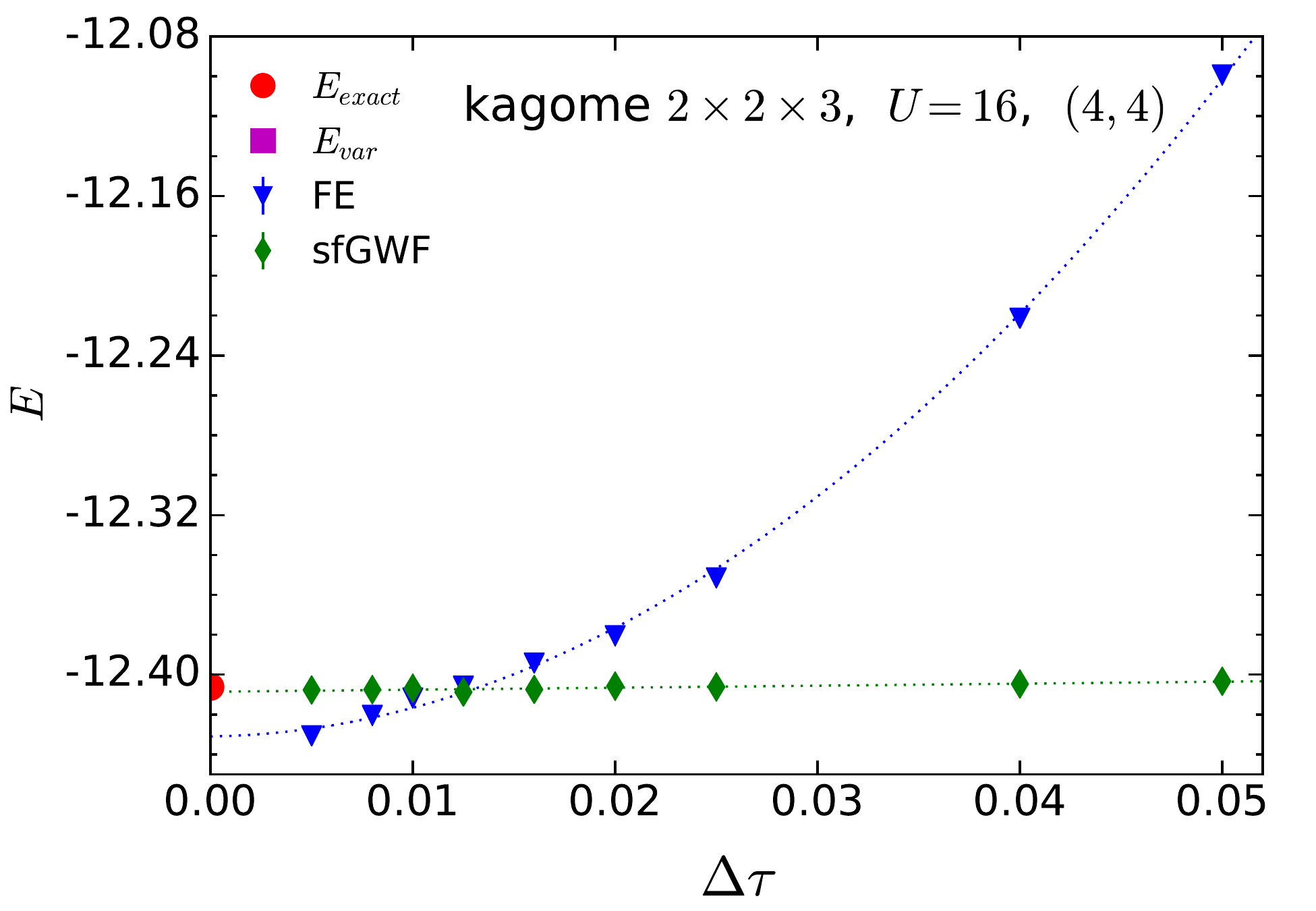}\hspace{4mm}
\includegraphics[scale=0.4]{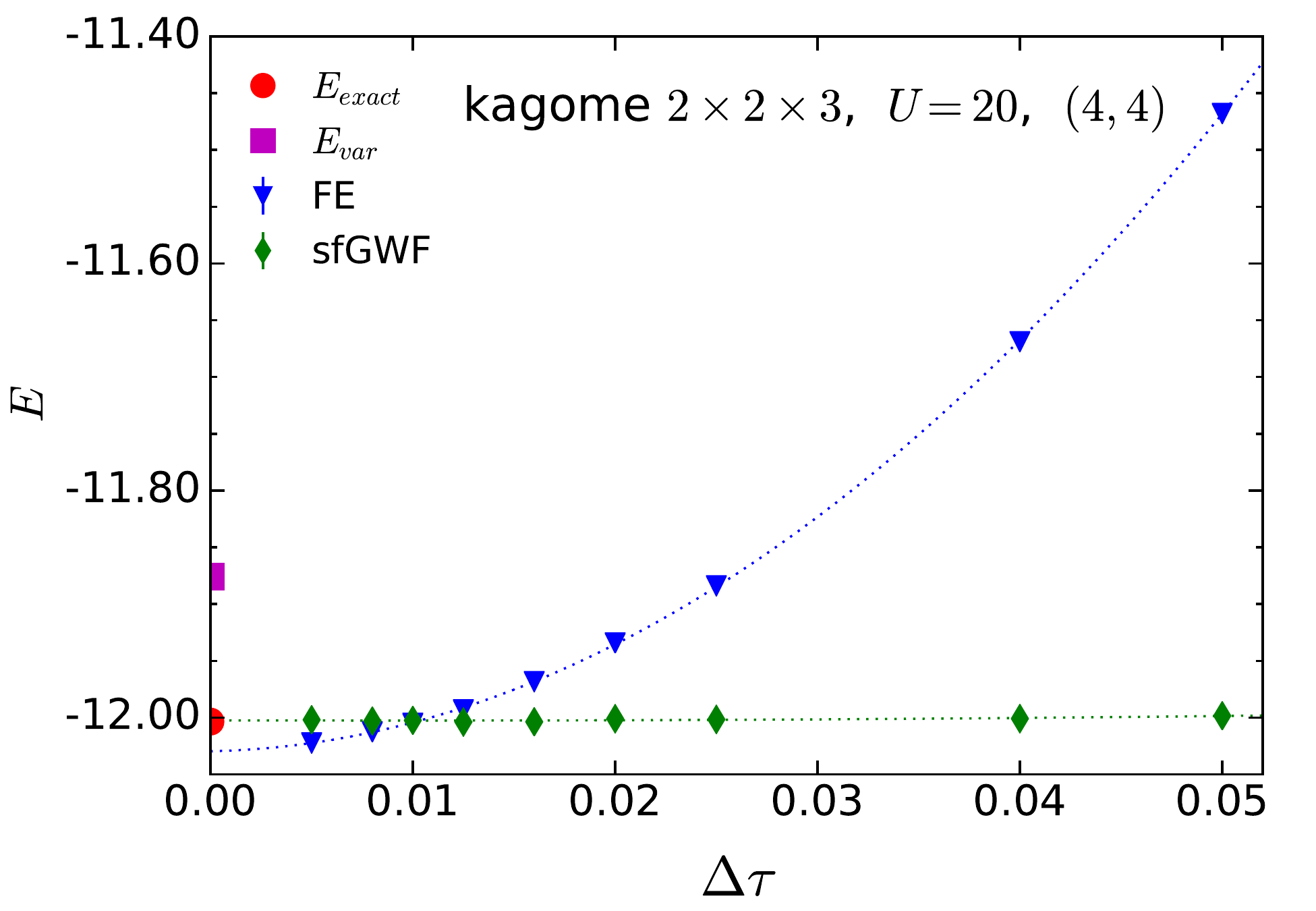}
\caption{Correction of the systematic Trotter approximation error for the 4-hole doped
$2\times 2\times 3$ kagome lattice. The geometry of the simulation cell is depicted in
the inset of Fig.~\ref{fig:CoeffKagome}.
The CPMC energy is plotted as a function of the time step $\Delta\tau$. The ED energy 
(red circle) and sfGWF variational energy (purple square) are also included for reference.
We note that the CPMC energy is not variational,\cite{Carlson:1999} as can be seen from 
the extrapolated results.
}
\label{fig:TrotterErrorKagome}
\end{figure*}

We list detailed sfGWF benchmark data for the square and kagome lattices in Table \ref{tbl:sfGWF}.
In Fig.~\ref{fig:TrotterErrorKagome}, Trotter corrections for the 4-hole doped $2\times 2\times 3$
kagome lattice are presented.
 
\begin{table*}
\begin{ruledtabular}
\begin{tabular}{@{}ccccccccc@{}}
           &                &  $U$  &  $E_{ex}$   &  $E_{var}$    & $E_{CPMC}^{sfGWF}$ &  $E_{CPMC}^{FE}$ & B.C.  & $\ket{\Phi}$  \\
  \hline
  $\sysa$  &  $(5,5)$       &   10  &  -4.28210   &   -3.94084    &  -4.2881(9)        &   -4.1285(31)    &  PBC  & FE   \\
           &                &   12  &  -3.68722   &   -3.32770    &  -3.6862(6)        &   -3.5316(12)    &       &      \\
           &                &   16  &  -2.87709   &   -2.53662    &  -2.8673(21)       &   -2.7259(17)    &       &      \\
           &                &   20  &  -2.35166   &   -2.04540    &  -2.3485(27)       &   -2.2037(17)    &       &      \\
  \hline
  $\sysc$  & $(8,8)$        &  10  &  -7.13238    &   -4.834(11)  &  -7.156(21)        &   -6.9214(38)    & TABC  & FE   \\
           &                &  12  &  -6.06247    &   -3.745(10)  &  -6.024(23)        &   -5.8942(43)    &       &      \\
           &                &  16  &  -4.64872    &   -2.582(7)   &  -4.641(10)        &   -4.4892(41)    &       &      \\
           &                &  20  &  -3.76123    &   -1.963(9)   &  -3.737(15)        &   -3.6114(29)    &       &      \\
  \hline
  $\sysc$  & $(7,7)$        &  10  & -11.11166    &   -9.742(10)  &  -11.2806(10)      &   -11.3393(24)   & TABC  & FE   \\ 
           &                &  12  & -10.28901    &   -8.839(10)  &  -10.4218(13)      &   -10.5095(17)   &       &      \\
           &                &  16  &  -9.23087    &   -7.766(9)   &   -9.2872(14)      &    -9.4049(31)   &       &      \\
           &                &  20  &  -8.58849    &   -7.174(10)  &   -8.5923(20)      &    -8.6963(36)   &       &      \\
  \hline
  $\sysf$  & $(4,4)$        &  10  &  -13.47310   &  -13.39881    &   -13.4750(1)      &  -13.4885(6)     & PBC   & FE   \\
           &                &  12  &  -13.02480   &  -12.93998    &   -12.0282(2)      &  -13.0464(8)     &       &      \\
           &                &  16  &  -12.40616   &  -12.28907    &   -12.4085(1)      &  -12.4309(9)     &       &      \\
           &                &  20  &  -12.00351   &  -11.85899    &   -12.0024(4)      &  -12.0296(11)    &       &      \\
\end{tabular}
\end{ruledtabular}
\caption{Ground state energy comparisons for the 2D Hubbard model. We have considered the square lattice 
(half-filled as well as 2-hole doped) and 4-hole doped $2\times 2\times 3$ kagome lattice. The second
column denotes the configuration of spin-up and spin-down electrons.
$E_{ex}$ denotes exact diagonalization results. $E_{var}$ is the variational energy of the sfGWF. 
The last two columns list the CPMC energies obtained using the sfGWF and free-electron (FE) trial 
wave functions, respectively. The column ``B.C.'' lists the boundary conditions implemented in the simulations.
The last column ``$\ket{\Phi}$'' gives the unprojected single-determinant state.
Note that the CPMC energy is not variational.\cite{Carlson:1999} However, it is possible to construct an
energy estimator that gives the upper bound of the ground state energy.\cite{Carlson:1999} Here we do not
address this issue.
}\label{tbl:sfGWF}
\end{table*}

\bibliography{reference}

\end{document}